\documentstyle[preprint,aps,epsfig]{revtex}

\newcommand{\MM}{\bbox{M}}

\newcommand{\hel}{${}^3$He}
\newcommand{\ttz}{$T_{20}$}

\begin{document}

\tightenlines 
\draft
\preprint{ISSN 1344-3879, ~RIKEN-AF-NP-334}

\title{
Reaction mechanism and characteristics of $T_{20}$ in d + ${}^3$He backward elastic scattering at intermediate energies}
\author{
M. Tanifuji${}^a$, S. Ishikawa${}^a$, Y. Iseri${}^b$, 
T. Uesaka${}^c$, N. Sakamoto${}^c$, Y. Satou${}^c$, K. Itoh${}^c$,
\\
H. Sakai${}^{d,c}$, A. Tamii${}^d$, T. Ohnishi${}^d$, K. Sekiguchi${}^d$, 
K. Yako${}^d$, S. Sakoda${}^d$, H. Okamura${}^e$,
\\
K. Suda${}^e$ and T. Wakasa${}^f$}
\address{${}^a$Department of Physics, Hosei University,
Fujimi, Chiyoda, Tokyo 102-8160, Japan}
\address{${}^b$Department of Physics, Chiba-Keizai College, 
Todoroki, Inage, Chiba 263-0021, Japan}
\address{${}^c$The Institute of Physical and Chemical Research (RIKEN), 
Saitama 351-0198, Japan}
\address{${}^d$Department of Physics, University of Tokyo, Hongo, Bunkyo, 
Tokyo 113-0033, Japan}
\address{${}^e$Department of Physics, Saitama University, Saitama 338-8570, Japan}
\address{${}^f$Research Center for Nuclear Physics, Osaka University, Osaka 567-0047, 
Japan} 
%

\maketitle

\begin{abstract}
For backward elastic scattering of deuterons by ${}^3$He, cross sections $\sigma$ and tensor analyzing power $T_{20}$ are measured at $E_d=140-270$ MeV.  
The data are analyzed by the PWIA and by the general formula which includes virtual excitations of other channels, with the assumption of the proton transfer from ${}^3$He to the deuteron. 
Using ${}^3$He wave functions calculated by the Faddeev equation, the PWIA describes global features of the experimental data, while the virtual excitation effects are important for quantitative fits to the $T_{20}$ data. 
Theoretical predictions on $T_{20}$, $K_y^y$ (polarization transfer coefficient) and $C_{yy}$ (spin correlation coefficient) are provided up to GeV energies.
\end{abstract}



\narrowtext

\section{Introduction}

%
For the last few decades, elastic scattering  of deuterons by protons at the backward angle at intermediate energies has intensively been studied as an important source of information of nuclear interactions and reaction dynamics, by including non-nucleonic degrees of freedom in the consideration \cite{Ba72,Va73,Ig72,Ke81,Ar84,Bo85,Ko86,Ko94,Pu95,Ta98}. 
Since \hel{} has the same spin as the proton's, the spin structure of the scattering amplitude of the d+${}^3$He (d${}^3$He) system is similar to that of the d+p (dp) one when \hel{} is considered as a single body \cite{Re98}.  
Then the backward elastic scattering of the deuteron by \hel{} attracts our attention, for investigations of probable differences as well as similarities of the information when compared to the dp scattering.

In the dp backward scattering, the assumption of mechanism by neutron transfer from the deuteron to the proton (Fig.\ \ref{fig1}(a)) \cite{Va73} has been fundamentally successful in explaining energy dependence of observables.  
For example, the neutron-transfer mechanism produces the overall agreement to the experimental cross section $\sigma$ of the scattering up to GeV-energy region when the empirical momentum distribution of the deuteron is employed in the relativistic framework \cite{Ko86} and  the simple PWIA calculation by the mechanism describes qualitative features of the measured tensor analyzing power $T_{20}$ and polarization transfer coefficient $K^y_y$(d$\to$p) at few hundreds MeV \cite{Ar84,Pu95}.  
Further, the measured $\sigma$ and \ttz{} of the scattering have broad resonance-like structures around $E_d$ = 1 GeV in the plot against the incident energy \cite{Ar84,Pu95}, which have been interpreted as the signal of excitations of the transferred neutron to $\Delta$-states (Fig.\ \ref{fig1}(b)) \cite{Ar84,Bo85}.
The PWIA calculation by the transfer mechanism has reproduced the observed structure of $T_{20}$ when effects of other reaction channels like the $\Delta$ excitation have been phenomenologically included \cite{Ta98}.  
For the transfer mechanism, the spin observables are described by the $w(k)$ to $u(k)$ ratio, where $u(k)$ and $w(k)$ are the deuteron $S$- and $D$- wave functions in the momentum space, and thus the information of the nuclear interaction obtained by the spin observables is generally of different nature from the one by the cross section \cite{Ko94}.

In the case of the \hel{} target, the possible reaction mechanism of the backward scattering of the deuteron will be proton transfer from \hel{} to the deuteron (Fig.\ \ref{fig1}(c)) as follows. The backward cross section of direct scattering, for example by potentials, is small compared to the forward one by several order of magnitude at intermediate energies.  
On the other hand, the cross section of the proton transfer reaction is large for the forward emission of \hel, i.e. the backward for the deuteron, as in usual pick-up reactions.
When the transfer mechanism is adopted, the analogy to the dp scattering suggests that the spin observables of the d\hel{} scattering in the PWIA are described by the $w(k)$ to $u(k)$ ratio, where $u(k)$ and $w(k)$ are now the wave functions of the p-d relative motion in \hel.
The wave functions, which are supplied by the Faddeev calculation,  have ambiguities due to the choice of interactions as the input \cite{Sa86}. 
The information of the $w(k)$ to $u(k)$ ratio obtained from the spin observables will be useful in solving such ambiguities. 

Further, the spin structure of the scattering amplitude for the d${}^3$He system is similar to that for the dp one even in the nucleon transfer mechanism because in both scattering the spin of the projectile is transformed in the same way, from one to one half, and similarly that of the target from one half to one.  
Thus, the spin observables can be described in similar forms in these scattering and therefore some of their characteristics in the dp scattering will also be observed in the d${}^3$He one.  These stimulate us to investigate the spin observables of the d${}^3$He scattering at the backward angle by theories as well as by  experiments, which have not been attempted so far.  In particular, $T_{20}$ will be investigated in detail because of its sensitivity to the wave functions. 
The theoretical prediction is performed up to GeV-energy region similarly to the dp case although the present measurements are limited to a few hundreds MeV. 
As the first step of investigations, we will try  to clarify  global features of $T_{20}$ in the scattering theoretically and to provide experimental evidence for the reaction mechanism.  
The interests of the investigation will be focused on (i) the validity of the assumption of the proton-transfer mechanism, (ii) effects due to the difference between the ${}^3$He wave function and the deuteron one and (iii) the overview of effects of coupling to other reaction channels.  
In the following, experimental details are described in Sec. II, the PWIA analyses are given in Sec. III, where in addition to the cross section and the analyzing power the polarization transfer coefficient and the spin correlation one are briefly discussed, and in Sec. IV the effects of other reaction channels are investigated. 
Section V is devoted to a summary, discussion and perspective.

\section{Experimental details and results}

%
The d${}^3$He experiment was carried out at RIKEN Accelerator Research
Facility for $\sigma$ and $T_{20}$. Polarized deuteron beam was
provided by the high-intensity polarized ion source \cite{Ok93}.
Three polarization modes (unpolarized and two tensor-polarized modes)
were cycled every 5 seconds.
The beam polarization was measured with a polarimeter based on the $d+p$
elastic scattering after acceleration to 140, 200 and 270 MeV. 
It was monitored continuously during a run and obtained to be typically 60--80\% of the ideal value. 
At $E_{d}=$200 and 270~MeV, detectors were placed at the angle where $T_{20}$ vanishes so that the polarimeter could work as a beam intensity monitor independent of beam polarization.

A cryogenic ${}^3$He gas target was bombarded by the polarized deuteron beams. 
The size of the target was 13~mm (20~mm) wide, 15~mm (20~mm) high and 20~mm (10~mm) thick in the case of 270~MeV (140, 200~MeV) measurement. 
Entrance and exit windows were 6~$\mu$m thick Havar foils. 
The gas pressure ($\sim 1$~atm) was measured with a Baratron gauge. 
Temperature of the target was monitored by using a diode thermo-sensor and found to be about 11~K throughout the experiment. 
The density of the ${\rm ^3He}$ gas was $6.6\times 10^{20}$~cm$^{-3}$.

Scattered ${}^3$He particles were momentum-analyzed in the magnetic spectrograph SMART \cite{Ic94} and detected by a multi-wire drift chamber and plastic scintillators placed at the focal plane.  
The total beam charge was measured with a Faraday cup placed in the first dipole magnet and the beam intensity was monitored by the polarimeter described above.

In the off-line analysis, data for $\theta_{\rm lab}\le 1.4^{\circ}$ were used. 
The background spectrum from the $(d,{\rm ^3He})$ reactions on Havar foils was subtracted to obtain the net yield for the d${\rm ^3He}$ events. 
Signal-to-noise ratio at the peak region was 4--20\% depending on the beam energy.
The tensor analyzing power $T_{20}$ was deduced from the ratios of yields for three polarization modes.

The measured cross sections and analyzing powers with statistical errors are given in Table \ref{table1}. 
The systematic errors of $\sigma$ and $T_{20}$ are $\pm 10\%$ and $\pm 2\%$, respectively. 
The former is mainly due to the uncertainties of both the beam intensity and the target thickness, whereas the latter is due to the uncertainty of the beam polarization.

\section{PWIA analyses}

%
The theoretical analyses are performed first by the PWIA.
The approximation is rather crude but is still useful to obtain the qualitative feature of the reaction at the intermediate energies  \cite{Va73,Ko94}.  
Analyses by extended formulas which include the effect of virtual excitations of other reaction channels will be presented in the next section.

The \hel{} wave function in the pd cluster configuration is given by
\begin{eqnarray}
\Psi_{He} &=& \sum_{\nu_d} 
  \left( \frac12 1 \nu_p \nu_d | \frac12 \nu_{He}\right)
  \chi_{\nu_p} \chi_{\nu_d} \frac{u(k)}{\sqrt{4\pi}}
\nonumber\\
 && + \sum_{\nu_d,m}
  \left( \frac12 1 \nu_p \nu_d | \frac32 \nu\right)
  \left( \frac32 2 \nu m | \frac12 \nu_{He}\right)
  \chi_{\nu_p} \chi_{\nu_d} Y_{2m}(\hat{\bbox{k}}) w(k) ,
\label{eq1}
\end{eqnarray}
where $\bbox{k}$ is the proton-deuteron relative momentum in \hel{}, $\chi_{\nu_p}$ and $\chi_{\nu_d}$ are the wave functions of the proton and the deuteron, and $\nu^{,}$s are the z-components of their spins.
In the PWIA for the d\hel{} scattering, the proton transfer is induced by proton-deuteron interactions as the neutron transfer by neutron-proton interactions in the dp scattering \cite{Ko94}.
We denote the proton-deuteron scattering amplitude at the momentum $k$ by $t(k)$, neglecting the spin dependence similarly to the dp case. 
Referring to the calculation of the dp scattering, we get non-vanishing independent T matrix elements $\langle \nu_{He}^\prime, \nu_d^\prime | \MM | \nu_{He}, \nu_d \rangle$ for the proton transfer in the d${}^3$He backward elastic scattering as
\begin{equation}
\langle \frac12,1 | \MM | \frac12,1 \rangle = 
  \frac{t(k)}{4\pi}
  \left\{
  \frac23 u(k)^2 + \frac{2\sqrt{2}}{3} u(k) w(k) + \frac13 w(k)^2
  \right\} ,
\label{eq2}
\end{equation}
\begin{equation}
\langle \frac12,0 | \MM | -\frac12,1 \rangle =
\langle -\frac12,1 | \MM | \frac12,0 \rangle = 
  \frac{t(k)}{4\pi}
  \left\{
  \frac{\sqrt{2}}{3}u(k)^2-\frac13 u(k) w(k) - \frac{\sqrt{2}}{3} w(k)^2
  \right\} ,
\label{eq3}
\end{equation}
\begin{equation}
\langle -\frac12,0 | \MM | -\frac12,0 \rangle = 
  \frac{t(k)}{4\pi}
  \left\{
  \frac13 u(k)^2 - \frac{2\sqrt{2}}{3} u(k) w(k) + \frac23 w(k)^2
  \right\} ,
\label{eq4}
\end{equation} 
where $k$ is related to the incident deuteron momentum $k_d$ as $k=\frac13 k_d$ \cite{Ko94}.
The matrix elements (\ref{eq2})-(\ref{eq4}) are equivalent to those in the dp backward scattering except for the factor $\frac23$, when $\nu_{He}$ is replaced by $\nu_p$ and $t(k)$ by the n-p scattering amplitude. Then the cross section $\sigma$ and the tensor analyzing power $T_{20}$ are given as 
\begin{equation}
\sigma=(\frac{\mu}{2\pi})^2 \frac13 (\frac{|t(k)|}{4\pi})^2 (u(k)^2+w(k)^2)^2 
\label{eq51}
\end{equation}
and
\begin{equation}
T_{20} = \frac{2\sqrt{2} r -r^2}{\sqrt{2}(1+r^2)}
\quad \mbox{with} \quad
r = \frac{w(k)}{u(k)} .
\label{eq5}
\end{equation}
Here $\mu$ is the reduced mass of the d${}^3$He system. 
Using Eqs.\ (\ref{eq2})-(\ref{eq4}), one can also calculate other polarization observables, which are equivalent to those in the dp scattering. 
For example, the polarization transfer coefficient $\kappa_0=\frac32K_y^y$(d$\to$${}^3$He) and the spin correlation coefficient $C_{yy}$ are given as
\begin{equation}
\kappa_0=\frac1{1+r^2} \left(1-\frac1{\sqrt2} r -r^2 \right)
\label{eq52}
\end{equation}
and
\begin{equation}
C_{yy}=\frac2{9(1+r^2)^2} 
\left( 1-\frac5{\sqrt2} r +3r^2 +\sqrt2 r^3 -2 r^4 \right) .
\label{eq53}
\end{equation}
The quantities $T_{20}$ and $\kappa_0$ satisfy the equation of a circle as in dp scattering, 
\begin{equation}
(T_{20} + \frac1{2 \sqrt2})^2 + {\kappa_0}^2 =\frac98.
\label{eq54}
\end{equation}
At the low-energy limit where $r=0$, $T_{20}=0$ and $\kappa_0=1$. 
With the increase of the incident energy, the magnitude of $r$ is increased and the point defined by a set of $\kappa_0$ and $T_{20}$ in the $\kappa_0-T_{20}$ plane moves along the above circle clockwise for $r<0$ and counterclockwise for $r>0$. 
As will be seen later, the former is the case of the dp scattering and the latter that of the d${}^3$He one.
More details are referred to Refs.\ \cite{Ko94,Pu95,Ta98}. 

For the Faddeev calculation of \hel{}, we will specify the nucleon-nucleon (NN) force to the AV14 potential \cite{Wi84} and the three-nucleon (3N) force to the 2$\pi$-exchange Brazil model \cite{Ro86}.
The Coulomb interaction is discarded and the ${}^3$He nucleus is approximated by the triton. 
These potentials give the binding energy of the three-nucleon system as 8.34 (7.68) MeV with (without) the 3N force \cite{Sa86}. 
In Fig.\ \ref{fig2}(a), the calculated $u(k)$ and $w(k)$ are shown as the functions of $k$ and are compared to those for the deuteron by the same NN force. 
The wave functions of ${}^3$He have similar $k$-dependence to those of the deuteron in a global view but with the opposite sign of $w(k)$ for most $k$. 
Due to the different sign of $w(k)$, the calculated $r$ of ${}^3$He has the opposite sign to that of the deuteron except at large $k$ as is shown in Fig.\ \ref{fig2}(b). 
This characteristic of $r$ is consistent with the result of Ref.\ \cite{EI90} that the asymptotic D-state to S-state ratio in ${}^3$He has the opposite sign to that in the deuteron. 
The quantity $r$ is infinite at the zero point of $u(k)$, $k=k_{0u}$, which is located at almost the same $k$ for ${}^3$He and the deuteron, $k_{0u}=0.40$ GeV/c.
The zero point of $w(k)$ for ${}^3$He, $k=k_{0w}$, is located at about $k_{0w}$=0.79 GeV/c (0.96 GeV/c) with (without) the 3N force.

The analysis of the measured cross section will not fully been performed because the cross section depends on the proton-deuteron scattering amplitude $t(k)$ as is seen in Eq.\ (\ref{eq51}), and reliable information of $t(k)$ is not available at the present. 
In Fig.\ \ref{fig3}, however, a crude estimation which assumes the $k$-dependence of $t(k)$ to be negligible for the relevant energies is presented. 
The calculation describes the global feature of the energy dependence of the measured cross section. 
This will encourage further investigations of the scattering by the assumption of the transfer mechanism. 

The characteristics of $r$ in Fig.\ \ref{fig2} are reflected to $T_{20}$ of the scattering as shown in Fig.\ \ref{fig4}(a), where comparisons will be made in two ways; one is the comparison of $T_{20}$ between the d${}^3$He scattering and the dp one and the other is that of $T_{20}$ between the calculated and the measured for the same scattering. 
Due to the opposite sign of $r$, in a small-$k$ region, the calculated $T_{20}$ for the d${}^3$He scattering has the opposite sign to that for the dp one. 
In the figure, the measured $T_{20}$ for the d${}^3$He scattering by the present experiment is shown together with that for the dp scattering in Ref.\ \cite{Pu95} and their signs are opposite to each other. 
Since the dp and d${}^3$He scattering in the transfer model are essentially (d,p) stripping and (d,${}^3$He) pickup reactions, although the incident energies are higher than the conventional stripping and pickup reactions, the above feature is compared to that of measured $T_{20}$ for (d,p) and (d,t) reactions at low energies, where $T_{20}$ of the former reactions has the opposite sign to that of the latters in a wide angular range \cite{EI90,Kn75}. 
This nature of the sign of the measured $T_{20}$ indicates that the proton-pickup mechanism is a reasonable model for the d${}^3$He backward scattering. 
In both of the d${}^3$He and dp scattering, the calculated $T_{20}$ at the small $k$ reproduces the qualitative feature of the $k$ dependence of the measured one, although the magnitudes of the calculated are larger than those of the measured.  
This qualitative agreement of the calculation to the experiment supports the assumption of the proton-transfer mechanism for the d${}^3$He scattering. 
In Fig.\ \ref{fig4}(b), as an example of calculations by other potentials, $T_{20}$ by the recently proposed  NijmII potential \cite{St94} is compared to that by the AV14 one. 
The calculated $T_{20}$ are very similar to each other up to $k \approx 0.6$ GeV/c and the difference due to the potential employed is seen at larger $k$. 
In the figure, the effect of the 3N force, which arises through high-momentum components of the ${}^3$He wave function, is shown for the AV14 potential. 
In large $k$ region, the contribution of the 3N force to \ttz{} becomes large and the second zero point of \ttz{} is shifted from $k$=0.96 GeV/c to $k$=0.79 GeV/c due to the 3N force.

In Figs.\ \ref{fig5}(a) and \ \ref{fig5}(b), the calculated $\kappa_0$ and $C_{yy}$ are shown for the AV14 and  NijmII potentials. 
The dependence of $\kappa_0$ and $C_{yy}$ on the choice of the potential is weak, up to about $k \approx 0.6$ GeV/c. 
The 3N force effect is shown in the figure but is less remarkable compared to that in $T_{20}$.

\section{Virtual excitation of other reaction channels}

%
Now we will extend the theoretical framework to a more general one so as to include effects of other reaction channels \cite{Ta98}.
Let us first transform the original $T$-matrix elements 
$\langle \nu_{He}^\prime, \nu_d^\prime | \MM | \nu_{He}, \nu_d \rangle$ 
into $U$, $T$ and $T^\prime$ as
\begin{equation}
U = \frac9{2\sqrt{2}} \langle \frac12,1 | \MM | \frac12,1 \rangle
   + 3 \langle \frac12,0 | \MM | -\frac12,1 \rangle
   + \frac{3}{\sqrt{2}} \langle -\frac12,0 | \MM | -\frac12,0 \rangle ,
\label{eq7}
\end{equation}
\begin{equation}
T = 2 \langle \frac12,1 | \MM | \frac12,1 \rangle
   -\sqrt{2}  \langle \frac12,0 | \MM | -\frac12,1 \rangle
   -2 \langle -\frac12,0 | \MM | -\frac12,0 \rangle ,
\label{eq8}
\end{equation}
\begin{equation}
T^\prime = \frac14 \langle \frac12,1 | \MM | \frac12,1 \rangle
   - \frac1{\sqrt{2}} \langle \frac12,0 | \MM | -\frac12,1 \rangle
   + \frac12 \langle -\frac12,0 | \MM | -\frac12,0 \rangle .
\label{eq9}
\end{equation}
Here, $U$, $T$ and $T^\prime$ are free from the PWIA in principle.  
It is shown by the invariant-amplitude method \cite{Ta98} that $U$ is the scalar amplitude in the spin space and describes the scattering by central interactions and $T$ and $T^\prime$, the second-rank tensor ones, describe the scattering by tensor interactions. 
In the PWIA limit, as will be seen later, contribution of the S-state of ${}^3$He ($u(k)^2$) is included in $U$, that of the interference between the S-state and the D-state ($u(k)w(k)$) in $T$ and that of the D-state ($w(k)^2$) is divided into two amplitudes,  $U$ and $T^\prime$, according to their tensorial character.
Secondly, we introduce the relative magnitudes and phases between $U$, $T$ and $T^\prime$ as
\begin{equation}
\frac{|T|}{|U|} \equiv R, \quad 
\frac{|T^\prime|}{|U|} \equiv R^\prime, \quad 
\theta_T -\theta_U \equiv \Theta, \quad 
\text{and} \quad
\theta_{T^\prime} -\theta_U \equiv \Theta^\prime .
\label{eq10}
\end{equation}
Then we get the general form of \ttz $\;$ in terms of $R, R^\prime, \Theta$ and $\Theta^\prime$ as 
\begin{equation}
T_{20} = \{2\sqrt2 R \cos \Theta -R^2 -32{R^\prime}^2 +12R{R^\prime} 
\cos ({\Theta}^\prime - \Theta)\}/N_R,
\label{eq11}
\end{equation}
\begin{equation}
N_R = \sqrt2 +2\sqrt2 R^2 +34\sqrt2 {R^\prime}^2 -4R^\prime 
\cos {\Theta}^\prime, 
\label{eq12}
\end{equation} 
which is exact except for the proton-transfer assumption. 
Since these formulas are the same as those in the dp case \cite{Ta98}, we will follow their analyses.

We will calculate $R$ and $R^\prime$ by the PWIA and treat $\Theta$ and $\Theta^\prime$ as the adjustable parameters. 
This treatment of $\Theta$ and $\Theta^\prime$ takes into account effects of virtual excitations of other reaction channels phenomenologically since the virtual excitations of the channel induces imaginary parts in the transition amplitudes to vary $\Theta$ and/or $\Theta^\prime$. 
These effects can be neglected in $R$ and $R^\prime$ since the neglect has not induced significant errors in the calculation of $T_{20}$ of the dp scattering except at very large $k$ \cite{Ta98}. 
The range of $\Theta$ and $\Theta^{\prime}$ will generally be $-\pi \le \Theta (\Theta^\prime) \le \pi$. 
In the following, $\Theta$ and $\Theta^{\prime}$ are treated to be $k$-independent for simplicity, although $\Theta$ and $\Theta^\prime$ vary with $k$ in principle. The amplitudes $U$, $T$ and $T^\prime$ in the PWIA limit are calculated as
\begin{equation}
U=3\sqrt2 (u(k)^2 +\frac14 w(k)^2)t(k),
\label{13}
\end{equation}
\begin{equation}
T=3\sqrt2 u(k)w(k)t(k)
\label{14}
\end{equation}
and
\begin{equation}
T^{\prime} =\frac34 w(k)^2 t(k).
\label{eq15}
\end{equation}
Then $R$ and $R^\prime$ are obtained as
\begin{equation}
R = \frac{4|r|}{4 + r^2}
\quad \mbox{and} \quad 
R^\prime = \frac{r^2}{{\sqrt2} (4 + r^2)}.
\label{eq16}
\end{equation} 
In addition, when we choose the PWIA limit for $\Theta$ and $\Theta^\prime$, i.e. $\Theta=0$ for $k<k_{0u}$ and $k>k_{0w}$ and $\pi$ for $k_{0u}<k<k_{0w}$, and $\Theta^{\prime}=0$, Eqs.\ (\ref{eq11}), (\ref{eq12}) and (\ref{eq16}) give the previous result, Eq.\ (\ref{eq5}). 
Since $\Theta$ changes at $k=k_{0u}$ and $k=k_{0w}$, in the following we will identify $\Theta$ by the value for $k<k_{0u}$, for simplicity.

By the numerical calculation, we will study the effects on $T_{20}$ by varying $\Theta$ and/or $\Theta^\prime$, specifying the potential to the AV14 one. 
First we will examine the calculation without the 3N force. 
In Fig.\ \ref{fig6}(a), effects of variations of $\Theta$ are described for typical $\Theta$ by fixing $\Theta^{\prime}$ to the PWIA limit, where the result is shown for positive $\Theta$ since $T_{20}$ in this case is independent of the sign of $\Theta$ as seen in Eqs.\ (\ref{eq11}) and (\ref{eq12}). 
The $k$ dependence of $T_{20}$ varies remarkably with $\Theta$. 
At $k=k_{0u} (r=\infty)$ and $k=k_{0w} (r=0)$, $T_{20}$ are independent of $\Theta$ and are $-\frac1{\sqrt2}$ and zero, respectively. 
The calculation by $\Theta=60^\circ$ describes the present data very well and indicates the importance of the virtual excitation of other channels for the quantitative description of the data. 
In Fig.\ \ref{fig6}(b), we show the $\Theta^\prime$ dependence of $T_{20}$ with typical $\Theta^\prime$ by fixing $\Theta=0$, where the result is shown for positive $\Theta^\prime$ because of the independence on the sign of $\Theta^\prime$ (see Eqs.\ (\ref{eq11}) and (\ref{eq12})). 
None of the calculations with variations of $\Theta^\prime$  reproduces our data of $T_{20}$ in the small-$k$ region. 
However, the calculations for $\Theta^\prime=120^\circ$ and $180^\circ$ produce resonance-like structures around $k=k_{0u}$, which are similar to the structure found in the dp scattering observed at $k=0.3-0.45$ GeV/c in Fig.\ \ref{fig4}(a).
There the calculation by Eqs.\ (\ref{eq11}) and (\ref{eq12}) with $\Theta=180^\circ$ (PWIA limit) and $\Theta^\prime=120^\circ$ for the dp scattering is shown to exhibit that the virtual-excitation effect reproduces the structure. 

As examples of  combined effects of $\Theta$ and $\Theta^\prime$, we vary $\Theta^\prime$ fixing $\Theta=60^\circ$ in Fig.\ \ref{fig6}(c). 
The calculated $T_{20}$ in the small-$k$ region is little affected by the variation of $\Theta^\prime$, reproducing our data. 
On the other hand, the calculations by $\Theta^\prime=-60^\circ$, and $ -120^\circ$ produce structures  similar to those in Fig.\ \ref{fig6}(b). 
Therefore $\Theta$ and $\Theta^\prime$  play different roles in the calculation of $T_{20}$; i.e. $T_{20}$ in the small-$k$ region is mainly governed by the magnitude of $\Theta$, while the resonance-like structure in the medium $k$ is produced by proper choices of $\Theta^\prime$. 
In Fig.\ \ref{fig6}(c), the calculated $T_{20}$ at $k=k_{0u}$ is concentrated in a narrow range of the magnitude of $T_{20}$. 
This is due to the weak dependence of $T_{20}$ on $\Theta^\prime$ at $k=k_{0u}$ as seen in Eqs.\ (\ref{eq11}) and (\ref{eq12}), and thus the measurement at $k=k_{0u}$ will provide less ambiguous examinations of the reaction mechanism. 
These features of $T_{20}$ are similar to those in the dp scattering \cite{Ta98}.

In Figs.\ \ref{fig6}(a) and \ \ref{fig6}(b), we show the effect of the 3N force, as examples, for $\Theta=120^\circ$ in Fig.\ \ref{fig6}(a), and $\Theta^\prime=120^\circ$ in Fig.\ \ref{fig6}(b). 
The qualitative nature of the effect is similar to that in the PWIA limit in Fig.\ \ref{fig4}(b); $T_{20}$ for large $k$ is remarkably affected by the 3N force, reflecting that the 3N force dominantly affects the p-d wave function at small distance. 
The second zero point of $T_{20}$, which is independent of $\Theta$ and $\Theta^\prime$, is moved to $k=0.79$ GeV/c from $0.96$ GeV/c by the force.

\section{Summary, discussion and perspective}

%
In the present paper, we have investigated the cross section and the spin observables in the d${}^3$He backward elastic scattering by the theory and the experiment. 
The theory which assumes the proton-transfer mechanism has predicted $T_{20}$, $\kappa_0$ and $C_{yy}$ in a wide range of the intermediate energies. 
In the low-energy region of the range, $\sigma$ and $T_{20}$ have been measured.
The global feature of the energy dependence of $\sigma$ is reproduced by the PWIA calculation with the energy-independent p-d scattering amplitude and the measured $T_{20}$ is described qualitatively by the PWIA prediction. 
The ${}^3$He wave function is obtained by solving the Faddeev equation. 
At small $k$, the sign of $w(k)/u(k)$ of ${}^3$He is different from that of the deuteron and this explains the sign of the measured $T_{20}$ in the d${}^3$He scattering to be opposite to that in the dp scattering.
These give strong support to the assumption that the dominant reaction mechanism is the proton transfer from ${}^3$He to the deuteron. 
The quantitative difference between the measured $T_{20}$ and the PWIA one has been explained by including the effect of the coupling to other channels. 
More details will be discussed later. 
The formula of $T_{20}$ in the d${}^3$He scattering looks similar to the one in the dp scattering. 
However, the difference of the internal wave functions between ${}^3$He and the deuteron produces the different features of $T_{20}$ for most $k$. 
The calculated $T_{20}$ depends weakly on the NN interaction employed, up to $k \approx 0.6$ GeV/c. 
The 3N force contributes to $T_{20}$ through the high-momentum components of the p-d wave functions in ${}^3$He. 
The 3N forces, which have a long history since the pioneer works in Ref. \cite{Fu57}, have recently been investigated in detail by analyzing cross sections, analyzing powers and some spin-correlation effects in the n-d scattering \cite{Wi99,Is99} and  the cross sections in the n-t one \cite{Ci99}. 
The measurement of $T_{20}$ at large $k$ in the d${}^3$He backward elastic scattering will provide additional information of the 3N force. 

\begin{figure}
\begin{center}
\epsfig{file=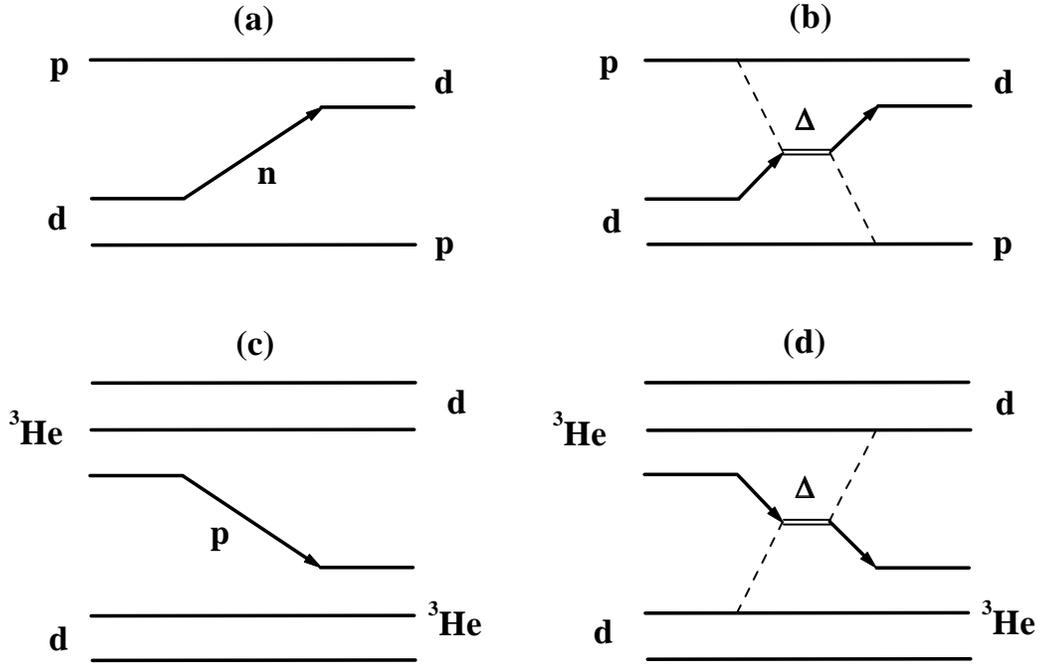, height=12cm}
\end{center}
\caption{
Neutron transfer in dp scattering and proton transfer in d${}^3$He scattering. 
(b) and (d) show the excitations of the transferred nucleons to $\Delta$-states by the exchange of mesons, for example pions, and include varieties of the combination of the charges of the related nucleons and those of the exchanged mesons.
In (d), the four combinations of the spectator nucleons  are described representatively by one combination. 
\label{fig1}}
\end{figure}

\begin{figure}
\begin{center}
\epsfig{file=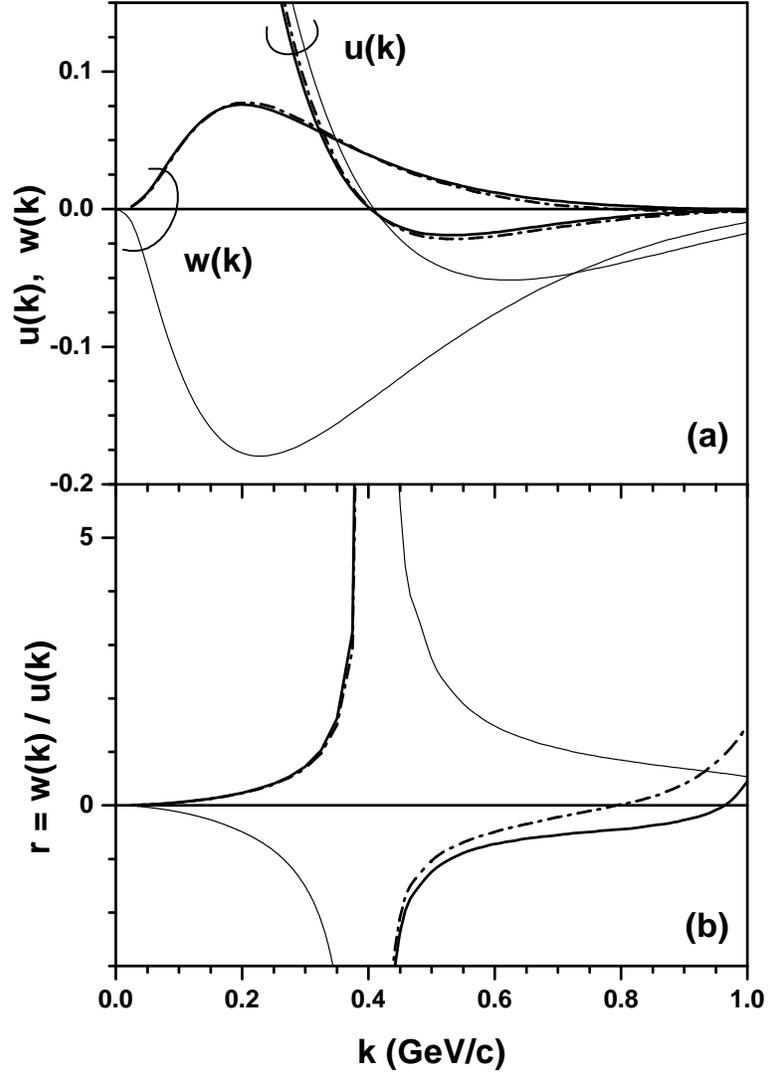, width=12cm}
\end{center}
\caption{
(a) Wave functions of ${}^3$He and deuterons normalized properly. 
$u(k)$ and $w(k)$ are the S and D components, respectively. 
(b) The D to S wave-function ratio $r$ for ${}^3$He and deuterons. 
The solid lines are for ${}^3$He and the thin solid lines for the deuteron. 
The dash-dotted lines are for ${}^3$He including the contribution of the Brazil 3N force. 
\label{fig2}}
\end{figure}

\begin{figure}
\begin{center}
\epsfig{file=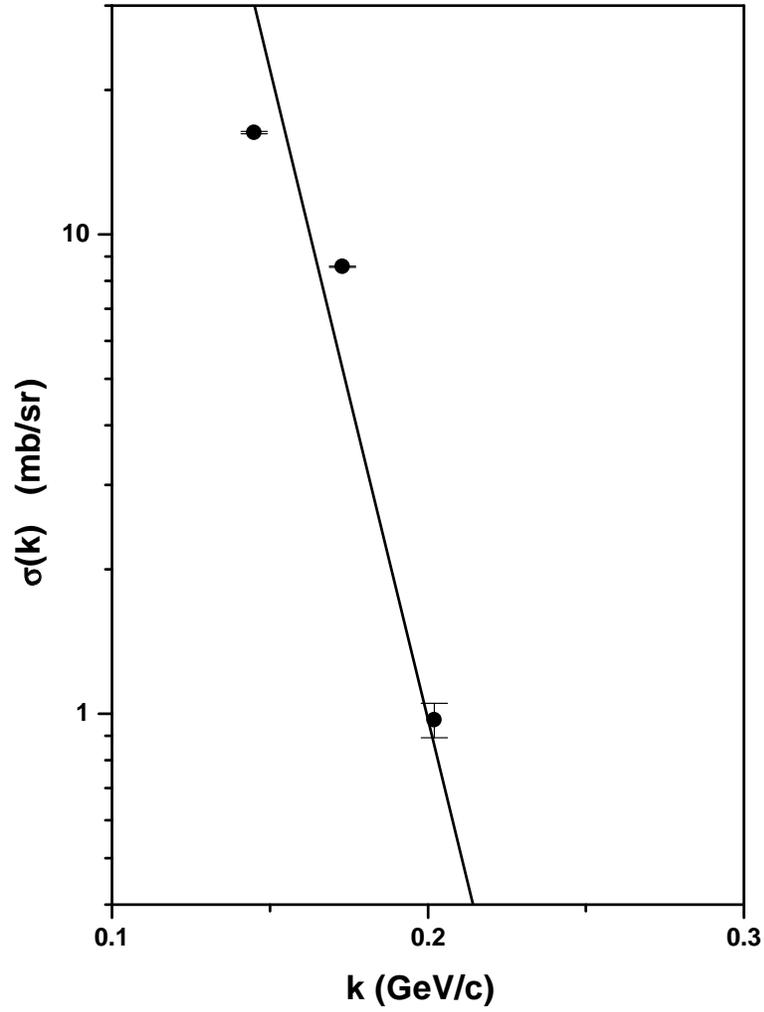, width=12cm}
\end{center}
\caption{
Cross sections of d${}^3$He backward elastic scattering. 
The closed circles are the present experimental data. 
The solid line is calculated by \protect{$\sigma \sim (u(k)^2+w(k)^2)^2$}, which is suitably normalized to the data.
\label{fig3}}
\end{figure}

The effect of the virtual excitation of other reaction channels is considered through the imaginary part of the scattering amplitudes, which are parameterized by $\Theta$ and $\Theta^\prime$. 
The calculations with $\Theta=60^\circ$ are successful in reproducing the small-$k$ data. 
This effect will be interpreted as the virtual-breakup effects, mainly of the deuteron, by the analogy to the dp case. 
In the case of dp scattering the calculations by $\Theta=120^\circ \sim 135^\circ$ which differ from the PWIA limit ($\Theta=180^\circ$) by $60^\circ \sim 45^\circ$ have described the measured $T_{20}$ at the small $k$ \cite{Ta98} and this $\Theta$ effect has been interpreted as the deuteron-breakup contribution, because below the pion threshold the breakup is the only one open channel strongly coupled to the dp one. 
In the present scattering, the large magnitude of $\Theta^\prime$ produces the resonance-like structure, which is similar to that observed in the dp scattering.
If the virtual excitation of the transferred nucleon to the $\Delta$-state (Fig.\ \ref{fig1}(b)) is responsible for producing the structure in the dp scattering, similar effects will also be expected in the d${}^3$He scattering as shown in Fig.\ \ref{fig1}(d). 
Thus it will be interesting to examine by experiments if such structures are observed. 
It should be noted that the effects of $\Theta$ and $\Theta^\prime$ on $T_{20}$ have been studied by using the AV14 NN potential and the Brazil 3N force. 
However, the essential features of the effects will be unchanged by other choices of the nuclear interactions except in the large-$k$ region. 

\begin{figure}
\begin{center}
\epsfig{file=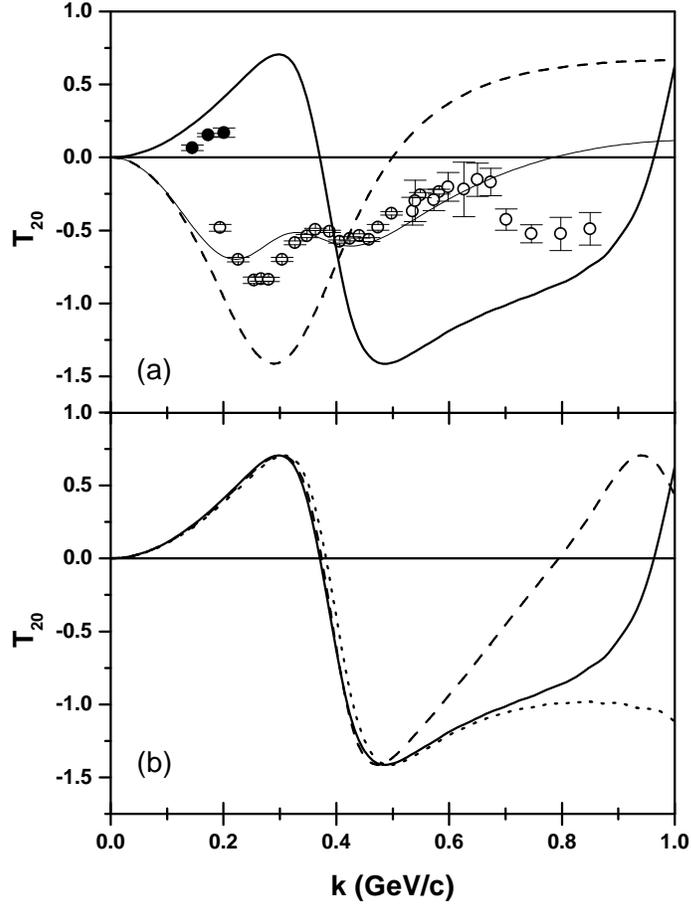, width=10cm}
\end{center}
\caption{
$T_{20}$ as a function of $k$. 
In (a), the solid and dashed lines describe the PWIA calculations by the AV14 potential for the d${}^3$He scattering and that for the dp scattering, respectively. 
The closed circles are the present experimental data for the d${}^3$He scattering and the open circles are the data for the dp scattering in Ref.\ \protect{\cite{Pu95}}. 
The thin solid line is for dp scattering, which includes effects of virtual excitations of other reaction channels. 
(b) describes the dependence of $T_{20}$ on the nuclear interactions employed for the d${}^3$He scattering. 
The solid and dotted line are for the AV14 and NijmII potentials, respectively and the dashed line includes the Brazil 3N force in the case of the AV14 potential.
\label{fig4}}
\end{figure}

\begin{figure}
\begin{center}
\epsfig{file=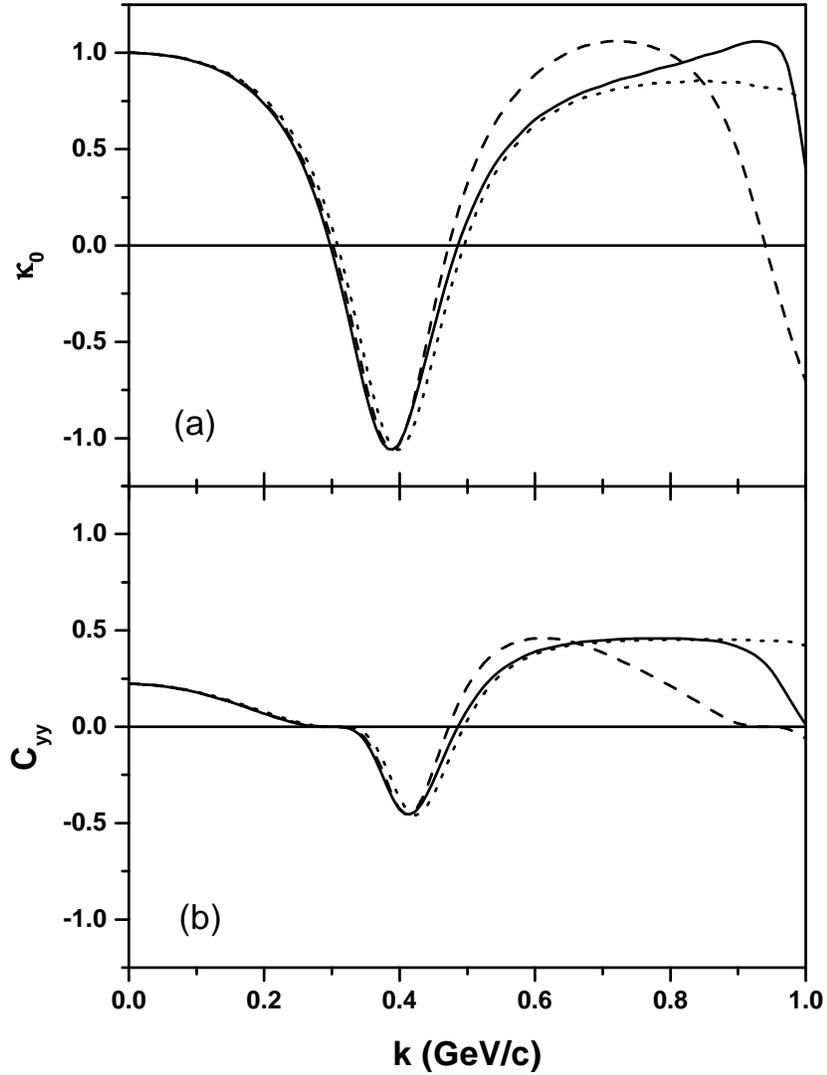, width=12cm}
\end{center}
\caption{
The PWIA calculations of $\kappa_0$ (a) and $C_{yy}$ (b) in d${}^3$He scattering as functions of $k$. 
The solid and dotted lines are for the AV14 and NijmII potentials, respectively and the dashed lines include the Brazil 3N force in the case of the AV14 potential.
\label{fig5}}
\end{figure}

\begin{figure}
\begin{center}
\epsfig{file=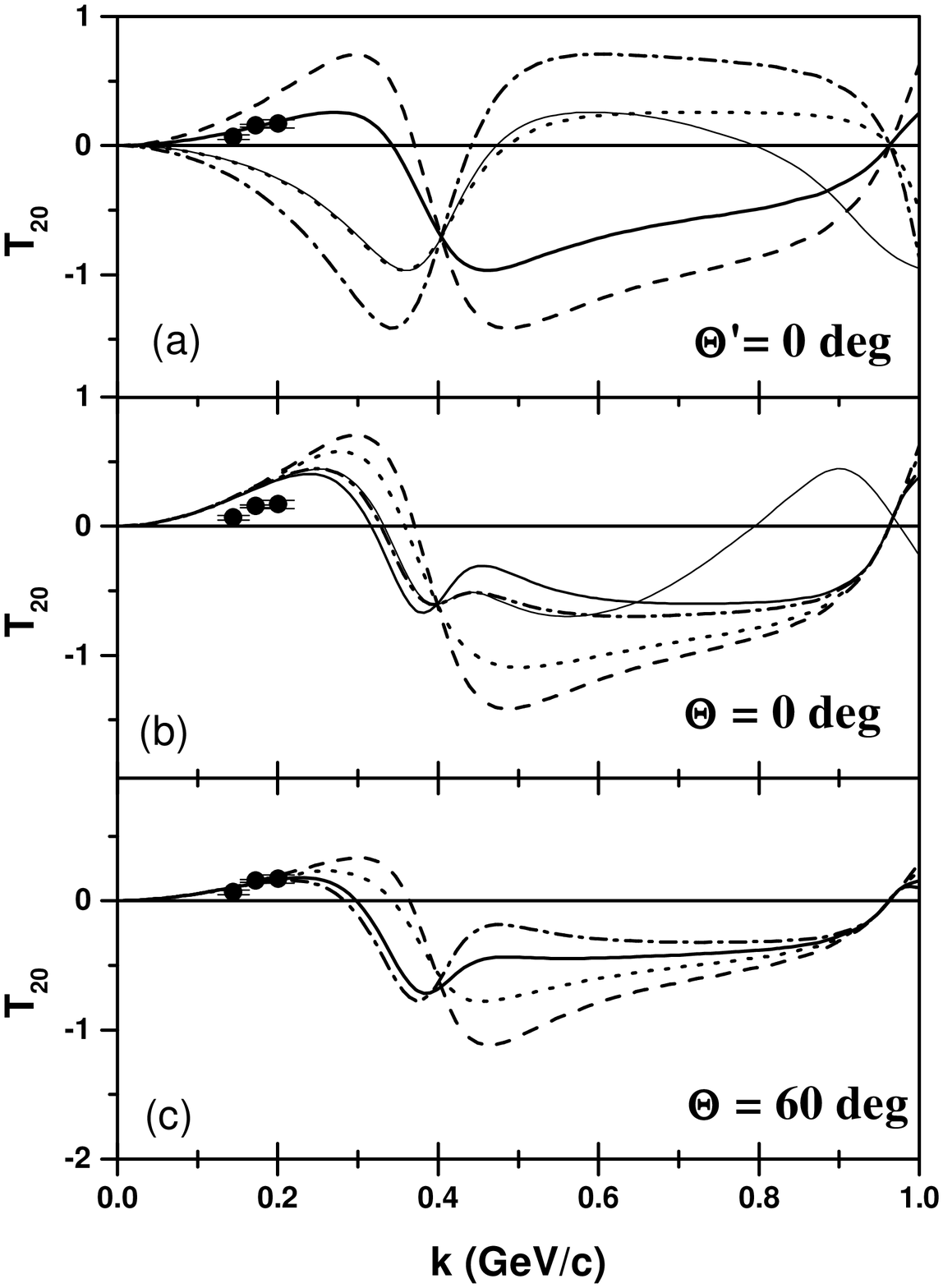, width=10cm}
\end{center}
\caption{
Effects of virtual excitations of reaction channels in d${}^3$He scattering.  
The calculations are by the AV14 nuclear potential. 
In (a), the dashed, solid, dotted and dash-dotted lines are the calculations for $\Theta=0^\circ, 60^\circ, 120^\circ$ and $180^\circ$ fixing $\Theta^\prime=0^\circ$ without the 3N force, respectively. 
The thin solid line includes the 3N-force effect for $\Theta=120^\circ$. 
In(b), the dashed, dotted, dash-dotted and solid lines are the calculations for $\Theta^\prime=0^\circ, 60^\circ, 120^\circ$ and $180^\circ$ fixing $\Theta=0^\circ$ without the 3N force, respectively. 
The thin solid line includes the 3N-force effect for $\Theta^\prime=120^\circ$.
In (c), the dashed, dotted, solid and dash-dotted lines are the calculations for $\Theta^\prime=60^\circ, 120^\circ, -60^\circ$ and $-120^\circ$ fixing $\Theta=60^\circ$ without the 3N force, respectively. 
The closed circles are the present experimental data for the d${}^3$He scattering. 
\label{fig6}}
\end{figure}

In the present investigation, the $k$ dependence of $\Theta$ and $\Theta^\prime$ is not considered. 
Since they depend on $k$ in principle, a particular set of $\Theta$ and $\Theta^\prime$ might be valid in a limited range of $k$. 
At present, $\Theta=60^\circ$ with any $\Theta^\prime$ reproduces the data between $k=0.1$ and $0.2$ GeV/c, indicating that such sets of $\Theta$ and $\Theta^\prime$ are valid in this range of $k$. 
In the dp scattering, as shown in Fig.\ \ref{fig4}(a), a set of $\Theta$ and $\Theta^\prime$ reproduces the global feature of the data in a wide range of $k$.
This suggests a set of $\Theta$ and $\Theta^\prime$ to be applicable in a wide range of $k$ in the d${}^3$He scattering as well.

Finally, although numerical results are not displayed, the effects of virtual excitation of other channels on $\kappa_0$ and $C_{yy}$ are small at small $k$. 
However, their contributions have appreciable magnitudes at $k \sim 0.2$ GeV/c.
Therefore, the measurements of these quantities in this $k$ region will give information of $\Theta$ and $\Theta^\prime$, such as the validity of the phenomenological value, $\Theta \approx 60^\circ$. 
Because the predicted $T_{20}$ has interesting features, experiments at higher energies are desirable, which will provide valuable information of the nuclear interaction and the reaction mechanism. 
In particular, the measurement of $T_{20}$ at $k \sim 0.4$ GeV/c will be one candidate to obtain the convincing evidence of the reaction mechanism because $T_{20}$ there is almost independent of $\Theta$ and $\Theta^\prime$. 
Further refinements of the theory, which include relativistic effects, will be made when higher-energy data become available.

%

%

\begin{table}
\caption{
Measured cross sections and tensor analyzing powers in d${}^3$He backward elastic scattering with statistical errors.
\label{table1}
}

\begin{tabular}{lccc}
$E_{Lab}$ (MeV) & 140 & 200& 270\\
\hline
$\sigma$ (mb/sr) & $16.3\pm0.1$ & $8.56\pm0.03$ & $0.97\pm0.08$ \\
$T_{20}$ & $0.07\pm0.02$ & $0.15\pm0.01$ & $0.17\pm0.03$ \\ 
\end{tabular}
\end{table}

%

\begin{references}

\bibitem{Ba72}
G. Barry, Ann. Phys. {\bf 73}, 482 (1972).

\bibitem{Va73}
S. S. Vasan, Phys. Rev. D {\bf 8}, 4092 (1973).

\bibitem{Ig72}
G. Igo et al., Nucl. Phys. {\bf A195}, 33 (1972); 
L. Dubal et al, Phys. Rev. D {\bf 9}, 597 (1974); 
B. E. Bonner et al., Phys. Rev. Lett. {\bf 39}, 1253 (1977).

\bibitem{Ke81}
B. D. Keister, Phys. Rev. C {\bf 24}, 2628 (1981); 
B. D. Keister and J. A. Tjon, Phys. Rev. C  {\bf 26}, 578 (1982); 
P. P. Rekalo and I. M. Sitnik, Phys. Lett. B {\bf 356}, 434 (1995); 
L. P. Kaptari et al., Phys. Rev. C {\bf 57}, 1097 (1998).

\bibitem{Ar84}
J. Arvieux et al., Nucl. Phys. {\bf A431}, 613 (1984) 
and references therein.

\bibitem{Bo85}
A. Boudard and M. Dillig, Phys. Rev. C {\bf 31}, 302 (1985); 
A. Nakamura and L. Satta, Nucl. Phys. {\bf A445}, 706 (1985).

\bibitem{Ko86}
A. P. Kobushkin, J. Phys. G {\bf 12}, 487 (1986).

\bibitem{Ko94}
A. P. Kobushkin et al., Phys. Rev. C {\bf 50}, 2627 (1994).

\bibitem{Pu95}
V. Punjabi et al., Phys. Lett. B {\bf 350}, 178 (1995); 
L. S. Azhgirey et al., Phys. Lett. B {\bf 391}, 22 (1997).

\bibitem{Ta98}
M. Tanifuji, S. Ishikawa and Y. Iseri, Phys. Rev. C {\bf 57}, 2493 (1998).

\bibitem{Re98}
M. P. Rekalo, N. M. Piskunov and I. M. Sitnik, 
Few-Body Syst. {\bf 23}, 187 (1998).

\bibitem{Sa86}
T. Sasakawa and S. Ishikawa, Few-Body Syst. {\bf 1}, 3 (1986);
S. Ishikawa and S. Sasakawa, Few-Body Syst. {\bf 1}, 143 (1986).

\bibitem{Ok93}
H. Okamura et al., AIP Conf. {\bf 293}, 84 (1993).

\bibitem{Ic94}
T. Ichihara et al., Nucl. Phys. {\bf A 569}, 287c (1994).

\bibitem{Wi84}
R. B. Wiringa, R. A. Smith and T. L. Ainsworth, 
Phys. Rev. C {\bf 29}, 1207 (1984).

\bibitem{Ro86}
M. R. Robilotta and H. T. Coelho, Nucl. Phys. {\bf A460}, 645 (1986).

\bibitem{EI90}
A. M. Eiro and F. D. Santos, J. Phys. G: Nucl. Part. Phys. {\bf 16}, 
1139 (1990) and references therein.

\bibitem{Kn75}
L. D. Knutson, Proc. of the Fourth Symposium on Polarization Phenomena 
in Nuclear Reactions ed W. Gruebler and V. Koenig 
(Basel; Birkhauser Verlag), 243 (1975).

\bibitem{St94}
V. G. J. Stock, R. A. M. Klomp, C. P. F. Terhaggen and 
J. J. de Swart, Phys. Rev. C {\bf 49}, 2950 (1994).

\bibitem{Fu57}
J. Fujita and H. Miyazawa, Prog. Theor. Phys. {\bf 17}, 360 (1957); 
J. Fujita, M. Kawai and M. Tanifuji, Nucl. Phys. {\bf 29}, 252 (1962).

\bibitem{Wi99}
H. Wita\l a, W. Gl\"okle, J. Golak, D. H\"uber, H. Kamada and 
A. Nogga, Phys. Lett. B {\bf 447}, 216 (1999).

\bibitem{Is99}
S. Ishikawa, Phys. Rev. C {\bf 59}, R1247 (1999).

\bibitem{Ci99}
F. Ciesielski, J. Carbonell and C. Gignoux, Phys. Lett. B {\bf 447}, 199 (1999).

\end{references}
\end{document}